# A Migration-Assisted Deep Learning Scheme for Imaging Defects Inside Cylindrical Structures via GPR —A Case Study for Tree Trunks

Jiwei Qian, *Student Member, IEEE*, Yee Hui Lee, *Senior Member, IEEE*, Kaixuan Cheng, *Student Member, IEEE*, Qiqi Dai, *Member, IEEE*, Arda Yalcinkaya, Mohamed Lokman Mohd Yusof, James Wang, and Abdulkadir C. Yucel, *Senior Member, IEEE*

*Abstract*— Ground-penetrating radar (GPR) has emerged as a prominent tool for imaging internal defects in cylindrical structures, such as columns, utility poles, and tree trunks. However, accurately reconstructing both the shape and permittivity of the defects inside cylindrical structures remains challenging due to complex wave scattering phenomena and the limited accuracy of the existing signal processing and deep learning techniques. To address these issues, this study proposes a migration-assisted deep learning scheme for reconstructing the shape and permittivity of defects within cylindrical structures. The proposed scheme involves three stages of GPR data processing. First, a dual-permittivity estimation network extracts the permittivity values of the defect and the cylindrical structure, the latter of which is estimated with the help of a novel structural similarity index measure-based autofocusing technique. Second, a modified Kirchhoff migration incorporating the extracted permittivity of the cylindrical structure maps the signals reflected from the defect to the imaging domain. Third, a shape reconstruction network processes the migrated image to recover the precise shape of the defect. The image of the interior defect is finally obtained by combining the reconstructed shape and extracted permittivity of the defect. The proposed scheme is validated using both synthetic and experimental data from a laboratory trunk model and real tree trunk samples. Comparative results show superior performance over existing deep learning methods, while generalization tests on live trees confirm its feasibility for in-field deployment. The underlying principle can further be applied to other circumferential GPR imaging scenarios. The code and database are available at: https://github.com/jwqian54/Migration-Assisted-DL.

*Index Terms*— Deep learning, ground-penetrating radar, permittivity reconstruction, migration-assisted scheme, tree defects' imaging

Manuscript received 28 June 2025. This work was supported by the Ministry of National Development through the Cities of Tomorrow (CoT) Research and Development Programme under Award COT-V4-2020-6. *(Corresponding author: Abdulkadir C. Yucel, Yee Hui Lee.)*

Jiwei Qian, Yee Hui Lee, Kaixuan Cheng, Qiqi Dai, Yalcinkaya Arda, and Abdulkadir C. Yucel are with the School of Electrical and Electronic Engineering, Nanyang Technological University, Singapore 639798 (e-mail: qian0069@e.ntu.edu.sg; eyhlee@ntu.edu.sg; chen1519@e.ntu.edu.sg; daiq0004@e.ntu.edu.sg; e240224@e.ntu.edu.sg; acyucel@ntu.edu.sg).

Mohamed Lokman Mohd Yusof, and James Wang are with the Centre for Urban Greenery and Ecology, National Parks Board, Singapore 259569 (e-mail: mohamed_lokman_mohd_yusof@nparks.gov.sg, james_wang@nparks.gov.sg ).

## I. INTRODUCTION

GPR has been widely used in the detection, imaging, and parameter estimation of subsurface objects in many applications, including road inspection [1], land mine detection [2], bridge analysis [3], and tree root mapping[4] [5]. All these applications require data acquisition along a flat or a quasi-flat surface of the host medium surrounding the target of interest. However, when detecting and imaging the abnormal regions/defects within the cylindrical structures such as the columns in the cultural heritage field [6] [7] and tree trunks [8] [9] [10], the traditional signal processing [11] [12] and imaging algorithms [13] [14] developed based on the half-space assumption of the host medium become ineffective and inaccurate. Moreover, the multiple reflections within the cylindrical structure interfere with the reflections from the defects, hindering the detection and imaging of internal defects [15]. To this end, this study is dedicated to developing signal processing and deep learning schemes for imaging defects inside cylindrical structures.

Among these cylindrical structures, tree trunks are particularly critical, as they support the tree's structural stability and enable the transport of water and nutrients between the roots and the canopy. The tree trunks are prone to developing internal defects, such as cavities and decay, which can compromise structural integrity and can lead to tree falls. To image the internal defects of tree trunks, microwave imaging techniques [16] [17] [18] [19] are often employed, but several limitations remain. First, conventional imaging algorithms cannot extract the permittivity of the internal defects, a key indicator of severity [20]. Second, the multiple reflections within the tree trunk and environmental noise introduce clutter that traditional signal processing techniques cannot fully suppress, thereby degrading the imaging quality [21] [22]. Third, migration-based imaging algorithms require repeated executions of the algorithm over a range of the equivalent electromagnetic wave velocities of the host medium , resulting in a time-consuming imaging process [23]. Additionally, conventional entropy-based autofocusing techniques (AFTs) [24] used in migration algorithms often yield the approximate location of the defect with poor shape accuracy. On the other hand, microwave tomography [25] [26] can reconstruct the shape and permittivity of internal defects. However, the ill-posed and nonlinear nature of the associated inverse scattering

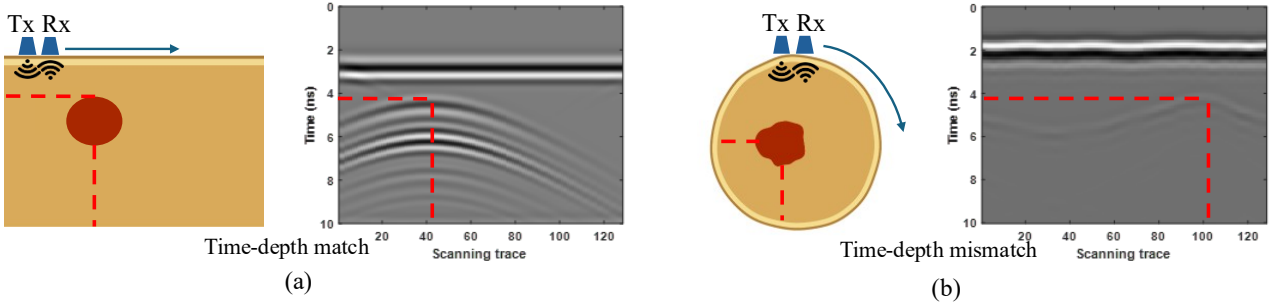

Fig. 1. Modelling and simulated B-scans of (a) flat surface scanning of a subsurface target and (b) circumferential scanning of the cylindrical structure with defect

problem necessitates the iterative use of electromagnetic forward solvers [27], making it impractical for real-time imaging of defects. To this end, there is a pressing need for a more robust approach capable of accurately reconstructing the shape and permittivity of defects inside tree trunks in near real-time.

Recently, deep learning schemes have shown promise in reconstructing shape and permittivity distributions in various GPR applications [28] [29] [30]. Among the prominent schemes, MRF-UNet [31] and PiNet [32] directly map B-scan inputs to permittivity profiles, whereas hybrid approaches, such as MultiPath-Net [33], integrate migrated images as auxiliary inputs. However, these methods are developed for flat-surface B-scan acquisitions and face critical challenges when applied to circumferential scanning of cylindrical structures. First, B-scan-based methods (e.g., MRF-UNet, PiNet) rely on the spatial alignment between reflections in the input B-scan and target locations in the imaging domain [34]. While flat-surface scanning produces hyperbolic patterns whose apex coincides with the target (see Fig. 1(a)), circumferential scanning yields sinusoidal reflections (see Fig. 1(b)) with no direct spatial correspondence, severely degrading model performance. Second, hybrid approaches also struggle under circumferential scanning due to spatial misalignment in B-scans and clutter in migrated images. Reflections from internal layers and bark surfaces introduce misleading scattering centers, complicating feature prioritization and weakening task comprehension. Moreover, migration algorithms are computationally expensive—requiring repeated trials over candidate wave velocities—and entropy-based AFTs often fail to produce accurate target shapes. Consequently, hybrid deep learning schemes relying on such suboptimal migrated images suffer significant performance degradation.

To address these challenges, this study proposes a migration-assisted deep learning framework specifically designed for reconstructing defects inside cylindrical structures. The proposed scheme first introduces a novel dual-permittivity estimation network (DPE-Net) to simultaneously predict the internal defect's permittivity and the host (cylindrical) structure's equivalent permittivity from the B-scan. To guide the selection of the host permittivity, a structural similarity index measure (SSIM)-based AFT is developed, which prioritizes preserving the defect's shape in the migrated image.

Then, using the estimated host permittivity, a modified Kirchhoff migration algorithm efficiently maps the reflected signals in the B-scan onto the boundary of the defect in the image domain with only a single execution, significantly reducing computation cost. Finally, a shape reconstruction neural network (SR-Net) refines the cluttered migrated image into a high-quality, clutter-free representation of the defect's shape. By combining the permittivity prediction from DPE-Net and the shape reconstruction from SR-Net, the proposed framework achieves near-real-time and accurate reconstruction of defects within cylindrical structures.

To the best of our knowledge, this work reports the first deployable and physics-consistent deep learning inversion framework tailored for GPR-based internal defect imaging in cylindrical objects. Its primary innovations and contributions are threefold:

1. A migration-assisted deep learning scheme is proposed to resolve the domain-mismatch bottleneck for cylindrical scanning while ensuring accurate permittivity estimation. The scheme enables physics-consistent interpretation through two complementary tasks: (i) permittivity estimation in the B-scan domain, where permittivity-dependent reflections are retained, and (ii) shape recovery in the migrated-image domain, where spatial geometry is correctly aligned. The embedded migration algorithm provides a reliable physics-guided transformation that aligns spatial information across domains, thereby effectively mitigating spatial misalignment.

2. A novel migration-integration strategy is developed for reliable defect imaging with a one-shot migration. The host structure's permittivity is learned by DPE-Net using supervisory labels generated via a novel SSIM-based AFT that prioritizes geometric fidelity in migrated images. This eliminates repetitive migration trials and significantly improves both efficiency and accuracy in downstream shape recovery.

3. Two domain-specialized networks are designed to maximize task performance. DPE-Net incorporates a feature aggregation module and a convolutional block attention module (CBAM) to enhance feature representation from B-scans, and SR-Net employs

ResPaths to bridge the semantic gap between encoder and decoder features, and an attention-based decoder to dynamically weigh the contributions of the features forwarded from different paths, further improving reconstruction performance.

Extensive experiments on simulation and measurement datasets and generalization tests on live trees demonstrate the feasibility and efficiency of the proposed scheme for practical tree inspections. It should be noted that our preliminary study [35] introduced SSIM-based autofocusing to improve the selection of migrated images for deep learning reconstruction in synthetic tree-trunk models. Nevertheless, the pipeline in [35] requires the true defect boundary during execution to determine the optimal equivalent permittivity of the host medium, which is not available in real-world tree inspection scenarios. In addition, the defect permittivity was inferred directly from the migrated images. However, migration inevitably weakens some material-dependent information originally contained in B-scan data (e.g. amplitude and time-of-arrival variations) which can reduce the accuracy of permittivity estimation. These limitations prevent [35] from being deployed beyond controlled simulation environments. To address these issues, this work presents a completely different reconstruction scheme to ensure robust permittivity map reconstruction along with a novel migration-integration strategy for automatic one-shot migration without any prior structural knowledge, resulting in a fully deployable, automated, and spatially consistent imaging system for real tree trunk inspections. Newly designed neural networks tailored to this paradigm, along with extensive measurement and field validations further confirm the accuracy, robustness, and practical utility of the proposed approach beyond simulation feasibility.

The rest of this paper is organized as follows. The methodology of the proposed scheme is explained in Section II. The performance of the scheme on the synthetic dataset of layered cylindrical objects, along with the ablation study and comparative study, are presented in Section III. The results of applications of the scheme to the measurement dataset of a laboratory tree trunk model are provided in Section IV. The accuracy of the proposed scheme on the measurement dataset of real tree trunk samples is expounded in Section V. The deployability of the proposed scheme in real-world applications is demonstrated in Section VI through generalization tests on live trees. Finally, the conclusions are drawn in Section VII. To unify the notations of the variables throughout the paper, we employ italic letters as the scalars (e.g., $a$ and $A$), boldface italic capital letters as the operators (e.g., $\boldsymbol{A}(\cdot)$), boldface lowercase letters as one-dimensional vectors (e.g., $\mathbf{a}$), boldface capital letters as two-dimensional matrices (e.g., $\mathbf{A}$), and boldface Euclid script letters as multi-dimensional tensors (e.g., $\mathcal{A}$).

## II. METHODOLOGY

The overall structure of the proposed scheme is illustrated in Fig. 2 and comprises three key components: dual-permittivity estimation, migration, and defect reconstruction. In the first stage, DPE-Net extracts and interprets key features from B-scans, and then estimates two permittivity values: the permittivity of the internal defect and the equivalent permittivity of the host cylindrical medium. The latter is obtained using the proposed SSIM-based AFT for training DPE-Net and serves as one of the inputs to the migration algorithm. In the second stage, the modified Kirchhoff migration algorithm is applied to map the signals reflected from the defect in the B-scan to the boundaries of the defect in the migrated image. In the final stage, the SR-Net further refines the migrated image by eliminating the clutter and revealing the accurate shape of the defect. By combining the predicted permittivity value from DPE-Net and the extracted shape of the defect from SR-Net, the permittivity map of the defect is reconstructed at the end. It should be noted that the SSIM-based AFT is used only during the offline training stage to generate supervisory labels for the DPE-Net. In practical field applications, the well-trained DPE-Net directly predicts the host-medium permittivity from the measured B-scan, without the need of executing the SSIM-based AFT. The details of each component are explained below.

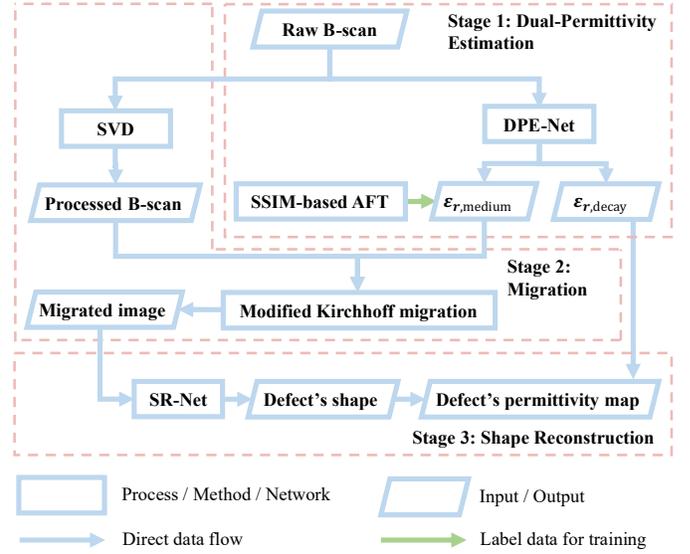

Fig. 2. Structure of proposed migration-assisted deep learning scheme

### A. Dual-Permittivity Estimation Network

The raw B-scan acquired during the circumferential scanning of a cylindrical structure is a 2-D radargram that contains the signals reflected from the layered host medium and the internal defect (see Fig. 1(b)). Typically, the reflections from the layers of the host medium are located between 1.5 and 3 ns, forming a (quasi-) parallel pattern in the raw B-scan [Fig. 1(b)], while the reflections from the defect are received within 4 and 6 ns, forming a sinusoidal pattern [Fig. 1(b)]. The DPE-Net is designed to analyze these two distinct features and extract the equivalent permittivity of the layered host medium, denoted as $\varepsilon_{r,\text{medium}}$, and the permittivity of the defect, denoted as $\varepsilon_{r,\text{defect}}$. $\varepsilon_{r,\text{medium}}$ serves as a critical input to the modified Kirchhoff migration algorithm in the next step of the proposed scheme. The architecture of the proposed DPE-Net is illustrated in Fig. 3 and consists of four main components: a feature extraction module (FEM), a feature aggregation module (FAM), a convolutional block attention module (CBAM), and a decoder.

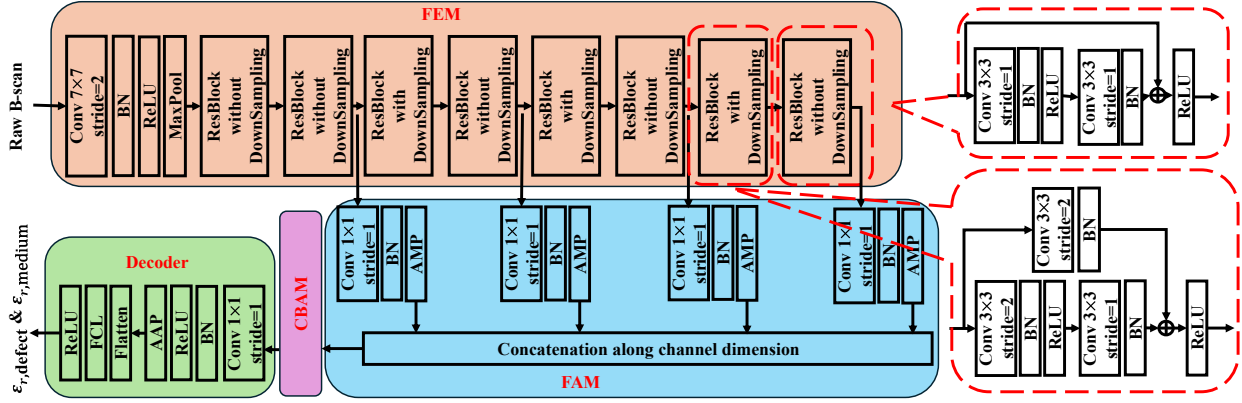

Fig. 3. The architecture of DPE-Net

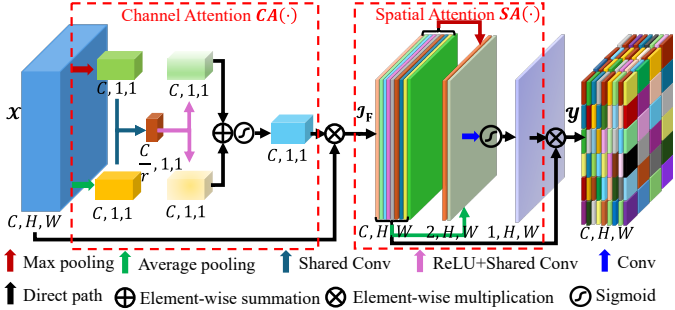

Fig. 4. Structure of CBAM

**FEM.** Inspired by the MLFF-Net architecture developed for tree defect detection [36], the DPE-Net utilizes the same feature extraction module as the backbone. The Resblock (residual block), as the fundamental feature extraction unit, is preferable due to its advantage of alleviating the vanishing gradient problem through skip connections. In DPE-Net [Fig. 3], eight Resblocks are cascaded to progressively extract the features of reflections associated with the layered host medium and internal defects and to encode from low levels to high levels. The resulting feature maps from each ResBlock capture a hierarchical understanding of how reflected signals relate to material properties, serving as a rich input for subsequent aggregation and attention modules.

**FAM.** Following feature extraction, the outputs of ResBlocks [Fig. 3] are passed to feature aggregation module to combine information across different levels of feature extraction. Specifically, each feature map is successively passed to a 1×1 2-D convolution (Conv) layer with a channel number 64 to add non-linearity for efficient representation of the feature, a batch normalization (BN) layer to stabilize training, and an adaptive max pooling (AMP) layer to adjust the dimensions of all feature maps to the same size of (4,4). The aggregation is finally completed by concatenating all feature maps along the channel dimension, producing an enhanced global representation of the features from the B-scan.

**CBAM.** To improve the discriminative power of the DPE-Net, a CBAM [37] is inserted between feature aggregation and decoder stages. CBAM refines the feature representation by sequentially applying the channel attention operator $CA(\cdot)$ and spatial attention operator $SA(\cdot)$. The structure of the CBAM is shown in Fig. 4. Let $\mathcal{X} \in \mathbb{R}^{C \times H \times W}$ denote the aggregated feature map, where $C$, $H$, and $W$ are its sizes along channel, height, and width dimensions, respectively. During channel attention stage [Fig. 4], CBAM reveals the inter-channel relationship of the feature map by applying average pooling and max pooling operations to retain the general and dominant features, respectively. The resulting two spatial context descriptors are then forwarded to two shared Conv layers and a rectified linear unit (ReLU) activation function with a reduced channel size of $C/r$ in between to reduce the computation cost. The outputs are combined by an element-wise summation, then passed through a sigmoid function, and element-wise multiplied by the original aggregated feature $\mathcal{X}$, yielding an ultimate channel-refined feature map $\mathcal{I}_F$.

During spatial attention stage [Fig. 4], $\mathcal{I}_F$ is subsequently passed to the spatial attention is evaluated in a complementary way. By average pooling and max pooling operations to squeeze the input feature map along the channel dimension, two individual channel context descriptors are generated and then concatenated and fed into a 7×7 Conv layer with a stride size of 1×1, and a padding size of 3×3 to evaluate the contribution of each element in the feature map. Followed by a sigmoid function, the spatial attention map is finally obtained. The final refined feature map $\mathcal{Y}$ is then calculated by multiplying :

$$\mathcal{Y} = \mathcal{X} \otimes CA(\mathcal{X}) \otimes SA(CA(\mathcal{X}) \otimes \mathcal{X}) \qquad (1)$$

where $\otimes$ is the element-wise multiplication.

**Decoder.** The attention-refined feature map is decoded by a Conv layer with a channel number 512 followed by a BN layer and a ReLU activation function [Fig. 3]. After adaptive average pooling (AAP), the decoded feature map that contains the comprehensive interpretation of the permittivity of interests is flattened into a vector and mapped into $\varepsilon_{r,medium}$ and $\varepsilon_{r,defect}$ via a fully connected layer (FCL) and a ReLU activation function simultaneously.

### B. Modified Kirchhoff Migration

The modified Kirchhoff migration has been developed to image internal defects in tree trunks [23], and is theoretically applicable for imaging the interiors of cylindrical structures with defects via circumference scanning. Unlike the full-wave inversion scheme, migration-based methods utilize the exploding source model (ESM), which treats the received signal as if it originated from the source at the defect location, to equivalently describe the two-way propagation of the EM wave in a real scenario. In particular, a couple of fictitious source points are assumed to "explode" at a reference time of $t = 0$ around the defect location. By radiating the EM wave from the exploded source points to the receiver, the two-way EM propagation and scattering process in the real scenario is approximately transferred to a one-way radiation problem. Therefore, the migration algorithm is introduced to estimate the intensity of the exploded source points from the B-scan.

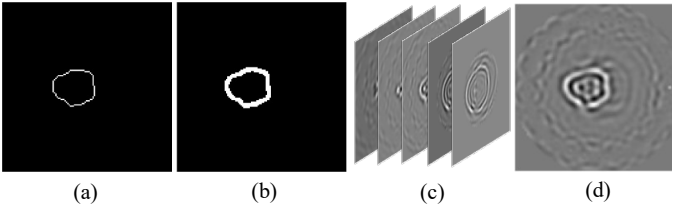

Fig. 5.(a) Detected boundary of defect, (b) ring mask, (c) candidate migrated images, and (d) selected migrated image with maximum SSIM

Considering the migration of a processed B-scan $\mathbf{G}(\mathbf{r}, t)$, the reconstructed image intensity $I$ at a specific location $\mathbf{r}_m = (x_m, y_m)$ can be obtained by the Kirchhoff integral along the 2-D enclosed outline of the cylindrical object [23]:

$$I(\mathbf{r}_m) = \frac{R_{min}}{2\pi} \int_0^L \frac{\partial R}{\partial n} \left( \frac{1}{R^2} \mathbf{G}\left(\mathbf{r}, \frac{R}{v}\right) - \frac{\mathbf{G}'\left(\mathbf{r}, \frac{R}{v}\right)}{vR} \right) ds \quad (2)$$

where $\mathbf{G}'(\mathbf{r}, t) = (\partial \mathbf{G}(\mathbf{r}, t) / \partial t)$ represents the derivative of the processed B-scan with respect to time $t$. Since the time-zero operation has been implemented on the processed radargram, $\mathbf{r} = (x, y)$ refers to the projected location of the cylindrical object from the scanning point on the trajectory, and the $L$ is the circumference of the cylindrical object. The distance between the investigated location and the point at the object's surface is represented by $R = \|\mathbf{r}_m - \mathbf{r}\|$, while the $(\partial R / \partial n)$ is the derivative of the distance $R$ in the direction normal to the integral contour. The scaling factor $R_{min}$ for a given investigation location is deployed to suppress the intensity bias for shallow investigated locations. Note that the solution validates based on the assumption that the medium between the internal defect and the cylindrical object surface is homogeneous, and the migration velocity $v$ is set to half of the actual electromagnetic wave velocity in the medium, following the one-way propagation assumption of the ESM [13].

### C. SSIM-based AFT

In the proposed approach, an SSIM-based AFT is introduced to determine the ground truth value of $\varepsilon_{r,medium}$ to train the DPE-Net. Unlike conventional entropy-based AFT, which maps the signals reflected from the defect in the B-scans to a focused zone, SSIM-based AFT emphasizes the accuracy of shape reconstruction of internal defects by evaluating the similarity between the migrated image and a ring mask that retains the boundary information of the internal defect. The complete AFT process consists of several steps: First, the edge of the internal defect [Fig. 5(a)] is captured by using the Sobel edge detector. Such a single-pixel detected boundary, as shown in Fig. 5(a), is further expanded to a ring mask [Fig. 5(b)] with a certain thickness. Second, a range of candidate $\varepsilon_{r,medium}$ is selected through empirical knowledge of the hosting medium. By executing the migration algorithm with each $\varepsilon_{r,medium}$, migrated images of all possible $\varepsilon_{r,medium}$ are generated [Fig. 5(c)]. Finally, the SSIM between each migrated image and the ring mask is evaluated via

$$SSIM(\hat{\mathbf{X}}_s, \mathbf{X}_s) = \frac{(2\mu_{\hat{\mathbf{X}}_s}\mu_{\hat{\mathbf{X}}_s} + c_1)(2\sigma_{\mathbf{X}_s\hat{\mathbf{X}}_s} + c_2)}{(\mu_{\hat{\mathbf{X}}_s}^2 + \mu_{\hat{\mathbf{X}}_s}^2 + c_1)(\sigma_{\hat{\mathbf{X}}_s}^2 + \sigma_{\hat{\mathbf{X}}_s}^2 + c_2)} \quad (3)$$

where $\mu_{\hat{\mathbf{x}}_s}$ and $\mu_{\mathbf{x}_s}$ are the means of the migrated image $\hat{\mathbf{X}}_s$ and the ring mask image $\mathbf{X}_s$, respectively. $\sigma_{\mathbf{X}_s}, \sigma_{\hat{\mathbf{X}}_s}$, and $\sigma_{\mathbf{X}_s\hat{\mathbf{X}}_s}$ are the variance of $\mathbf{X}_s$, variance of $\hat{\mathbf{X}}_s$, and covariance of $\mathbf{X}_s$ and $\hat{\mathbf{X}}_s$, respectively. $c_1$ and $c_2$ are two variables. The permittivity value of the migrated image that yields the maximum SSIM [Fig. 5(d)] is assigned as the $\varepsilon_{r,medium}$ for the host medium.

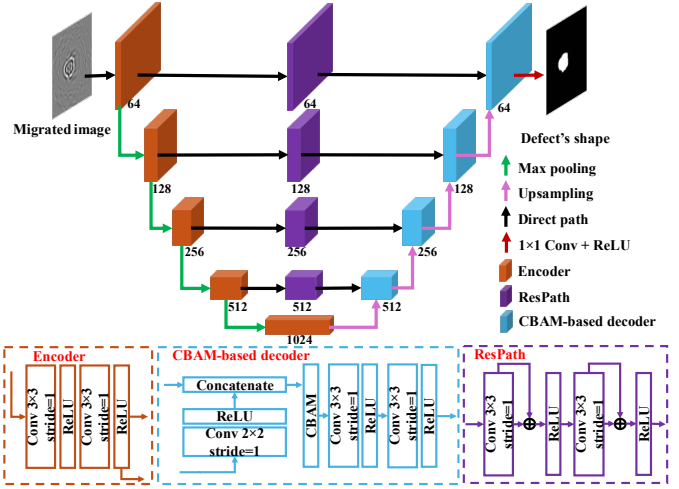

Fig. 6. The architecture of SR-Net

### D. Shape Reconstruction Network

Once the noisy migrated image is generated, the SR-Net takes it as the input, extracts the key features of the shapes, sizes, and locations of the internal defects, suppresses the clutter in the migrated images, and finally reconstructs the region of the defect accurately. The architecture of the SR-Net [Fig. 6] is optimized based on the conventional U-Net. It consists of the same encoder module, but different shortcut connections, named ResPath [38], between each level of encoder and decoder, and a CBAM-based decoder.

To be specific, the encoder is made up of five repeated applications of two 3×3 Conv layers and one max pooling layer with a stride of 2×2 to extract and compress the key features from the noisy migrated image, in which each Conv layer is followed by a ReLU function to increase the nonlinearity of the model. The consecutive operations of feature extraction enable the model to effectively encode the key features in the migration images from low to high levels.

During the decoding process, an upsampling of the feature map followed by a Conv layer and a ReLU is first applied to the feature map to increase its spatial size and halve the channel number. The generated feature map is then concatenated with the feature passed by ResPath between the same level of the encoder and decoder, enriching the global representation of the features of interest. Before forwarding the concatenated feature map to a decoder module that consists of two 3×3 Conv layers followed by two ReLU functions, the same CBAM [37] introduced in the DPE-Net is applied to refine the feature maps concatenated from two different paths. The output of the last decoder is mapped to the defect's geometry through 1×1 Conv layer with a ReLU activation function.

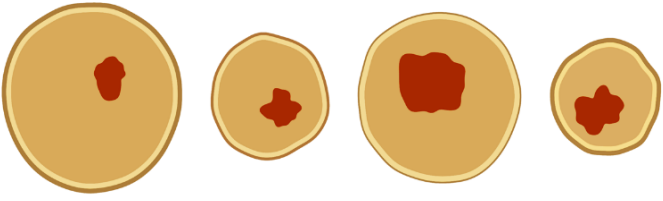

Fig. 7. Examples of synthetic cylindrical objects with internal defect

Instead of using the skip path as the shortcut connection in the conventional U-Net, the proposed SR-Net utilizes the ResPath [38] between each level of the encoder and decoder to resolve the following issues. 1) The semantic gap is observed in the concatenation of the same level of feature maps forwarded by the encoder $\mathcal{F}_{\text{encoder}}$ and the decoder $\mathcal{F}_{\text{decoder}}$. Specifically, $\mathcal{F}_{\text{encoder}}$ is supposed to be a low-level feature map as it is computed in the early layers of the network, while $\mathcal{F}_{\text{decoder}}$ is supposed to be a much higher level of feature map because of processing throughout more convolution layers. Merging these two incompatible sets of features directly could result in discrepancies throughout the learning process, thereby degrading the network's performance. 2) As the input, the migrated image selected by SSIM-based AFT contains defect information in the form of a ring, while the output of the network, which is the geometry of the defect, exists in the form of an area. With a flawed skip connection, the efficiency of image-to-image conversion between these two forms is limited. To this end, a ResPath is introduced to bridge between each level of encoder and decoder. By adding two pairs of one 3×3 Conv layer with the residual connection before the ReLU activation function, the features generated by the encoder are further adjusted to improve the performance of the network.

## III. NUMERICAL EXPERIMENT

### A. Numerical Dataset Preparation

To evaluate the performance of the proposed scheme, a dataset of synthetic B-scans of 3000 layered 2D cylindrical objects with internal defects is generated using a finite-difference time-domain method-based open-source electromagnetic solver, gprMax [39]. As shown in Fig. 7, both the cylindrical object and the internal defect are arbitrarily shaped to increase the diversity of the dataset. The radius of the cylindrical object and its internal defect vary within [15, 33]cm and [4, 23]cm, respectively. The thickness of the three layers of the object (from outer to inner) varies within [1, 2] cm, [1, 3] cm, and [12, 28] cm, respectively. Additionally, the defect is randomly located within the cylindrical objects. The relative permittivity/conductivity values of three layers of the object are randomly selected from the ranges of [2, 3]/[0.01, 0.03], [23, 25]/[0.1, 0.3], and [4, 5]/[0.01, 0.03], respectively. Moreover, the relative permittivity of the internal defect varies within [5, 40].

To generate the B-scan of the cylindrical object, a pair of transmitting (Tx) and receiving (Rx) antennas moves along a circular trajectory at a minimum distance of 5 cm from the object. A time-domain A-scan signal is recorded at every six-degree movement of the Tx-Rx pair, resulting in a B-scan composed of sixty A-scans. The transmitted signal is a Ricker waveform centered at 1 GHz. The input to the DPE-Net is the B-scan preprocessed by removing the direct coupling between the Tx and Rx. Once this coupling is removed, signatures corresponding to internal layered media and defects within the object become visible in the processed B-scan. This enhanced visibility facilitates more accurate feature extraction and prediction of $\varepsilon_{r,\text{medium}}$ and $\varepsilon_{r,\text{defect}}$.

To determine the ground truth of $\varepsilon_{r,\text{medium}}$, the modified Kirchhoff migration algorithm is executed with $\varepsilon_{r,\text{medium}}$ over the range of [2.5, 10] with a step of 0.5, and the SSIM-based AFT is applied to select $\varepsilon_{r,\text{medium}}$ via which a migration image with the maximum defect shape information is retained. To generate the output image of the SR-Net, the defect region is set to 1, while the rest of the region is set to 0.

### B. Implementation Details

The generated dataset is divided into training and testing datasets with an 80/20 split. The migrated image for the SR-Net is obtained by applying the predicted $\varepsilon_{r,\text{medium}}$ from the DPE-Net in the same dataset. All input and output images, as well as the predicted permittivity values, are normalized to the range [0, 1] based on the global maximum and minimum values across the entire dataset, while the sizes of all images are adjusted to 128×128. The mean square error (MSE) is used as the loss function and adjusted to the DPE-Net $loss_{\text{DPE}}$ and the SR-Net $loss_{\text{SR}}$ as

$$loss_{\text{DPE}} = |\mathbf{y}_{\text{defect}} - \hat{\mathbf{y}}_{\text{defect}}|^2 + |\mathbf{y}_{\text{medium}} - \hat{\mathbf{y}}_{\text{medium}}|^2 \quad (4)$$

$$loss_{\text{SR}} = \frac{1}{H \times W} \sum_{i,j} \left| \mathbf{Y}_{s_{i,j}} - \hat{\mathbf{Y}}_{s_{i,j}} \right|^2 \quad (5)$$

where $\mathbf{y}_{\text{defect}}$ and $\hat{\mathbf{y}}_{\text{defect}}$ are the ground truth and the prediction of $\varepsilon_{r,\text{defect}}$, respectively, and $\mathbf{y}_{\text{medium}}$ and $\hat{\mathbf{y}}_{\text{medium}}$ are the ground truth and the prediction of $\varepsilon_{r,\text{medium}}$, respectively. $\mathbf{Y}_s$ and $\hat{\mathbf{Y}}_s$ are the ground truth and the predicted geometry map of the defect, while the $i, j, H$, and $W$ are the indices and dimensions of the reconstructed geometry map by the SR-Net.

The networks are implemented through PyTorch and trained on an NVIDIA RTX 6000 GPU with a learning rate of 1e-4 and a batch size of 32 for 100 epochs, during which models with the lowest testing loss are saved. To evaluate the performance of the proposed scheme, several metrics are deployed in this study, including the mean relative error (MRE), mean absolute error (MAE) for the DPE-Net, and MSE, SSIM, and intersection over union (IoU) for the geometry map of the SR-Net and the permittivity map of the proposed scheme. The formulas are listed as follows:

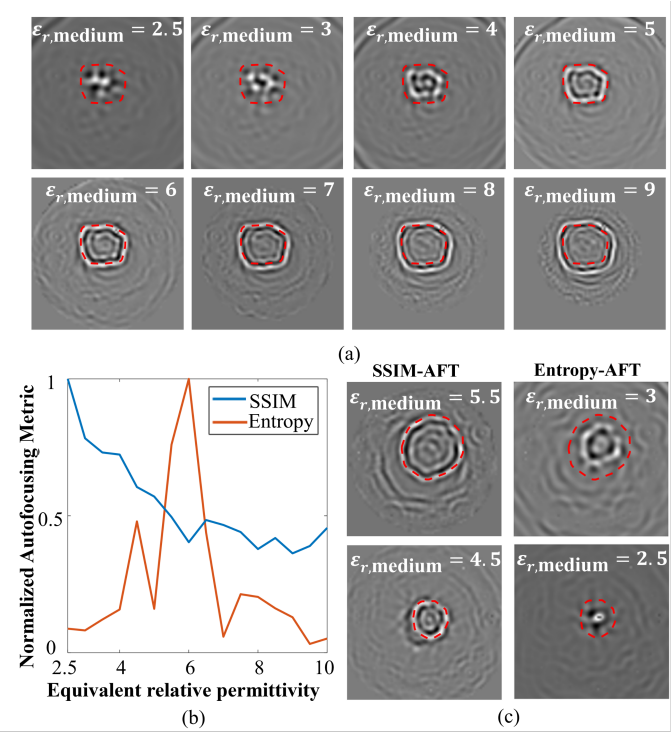

(a)

(b)

(c)

Fig. 8.(a) Migrated images under varying $\varepsilon_{r,\text{medium}}$, (b) corresponding normalized autofocusing metrics, and (c) comparison between entropy- and SSIM-based AFTs on two additional samples. Red dashed lines indicate the ground-truth defect outlines.

$$MAE_\alpha = |\mathbf{y}_\alpha - \hat{\mathbf{y}}_\alpha| \tag{6}$$

$$MRE_\alpha = \frac{|\mathbf{y}_\alpha - \hat{\mathbf{y}}_\alpha|}{\mathbf{y}_\alpha} \tag{7}$$

$$IoU = \frac{S_{\hat{\mathbf{Y}}_s \cap \mathbf{Y}_s}}{S_{\hat{\mathbf{Y}}_s \cup \mathbf{Y}_s}} \tag{8}$$

where $\alpha$ refers to either defect or medium, $S_{\hat{\mathbf{Y}}_s \cap \mathbf{Y}_s}$ and $S_{\hat{\mathbf{Y}}_s \cup \mathbf{Y}_s}$ represent the areas of the intersection and the union of the predicted shape and ground truth shape, respectively. The SSIM and MSE are evaluated using Equation (3) with the input of $(\hat{\mathbf{Y}}_s, \mathbf{Y}_s)$ and Equation (5), respectively.

### C. Effectiveness of SSIM-based AFT

To illustrate the distinct behaviors of the entropy-based [23] and SSIM-based AFTs, Fig. 8(a) presents migrated images generated from the same B-scan using different $\varepsilon_{r,\text{medium}}$, while Fig. 8(b) shows the corresponding normalized autofocusing metrics. As $\varepsilon_{r,\text{medium}}$ decreases, the migrated reflections become increasingly compact and tightly focused, whereas higher $\varepsilon_{r,\text{medium}}$ lead to more diffused and spatially spread patterns. Because the entropy-based AFT quantifies the overall energy concentration in the migrated image, it naturally assigns higher scores to these strongly focused cases and therefore tends to favor smaller permittivity values ($\varepsilon_{r,\text{medium}} = 2.5$). However, such excessive focusing compresses the reflection energy, producing images that appear visually sharper but are geometrically distorted. In contrast, the proposed SSIM-based AFT evaluates the structural similarity between the migrated image and the expected defect boundary. Accordingly, the selected $\varepsilon_{r,\text{medium}} = 6$ corresponding to the maximum SSIM value yields a migrated image that best preserves the defect's geometric structure. Moreover, as shown in Fig. 8(c), the SSIM-based AFT exhibits consistent focusing performance across samples with diverse defect sizes, shapes, and orientations, confirming its robustness and overall superiority.

For quantitative comparison, a root-mean-square (RMS)–equivalent permittivity $\varepsilon_{r,\text{RMS}}$ is adopted as a theoretical reference. The value is derived from the classical RMS velocity $v_{\text{RMS}}$ provided in [40], commonly used in seismic and GPR migration to approximate the kinematic behavior of layered media with a single effective velocity and improve migration quality [41]. Mathematically, $v_{\text{RMS}}$ and the corresponding $\varepsilon_{r,\text{RMS}}$ is computed as

$$v_{\text{RMS}}^2 = \frac{\sum_i v_i^2 t_i}{\sum_i t_i} \tag{9}$$

$$\varepsilon_{r,\text{RMS}} = \frac{c^2}{v_{\text{RMS}}^2} \tag{10}$$

where $v_i$ and $t_i$ denote the electromagnetic wave velocity and two-way travel time in $i$-th layered medium, respectively. To the best of our knowledge, no exact analytical expression exists that can yield a unique ground-truth effective permittivity for a cylindrical multilayered host medium under the migration-imaging framework. The $\varepsilon_{r,\text{RMS}}$ is employed not as a rigorous ground truth, but as a theoretically interpretable approximation that serves as a reference baseline for intuitively assessing whether the $\varepsilon_{r,\text{medium}}$ selected by different AFTs fall within a physically reasonable range.

Using the $\varepsilon_{r,\text{RMS}}$ as a benchmark, a statistical comparison is conducted between the proposed SSIM-based AFT and the conventional entropy-based AFT on the synthetic dataset. Quantitatively, the SSIM-based AFT achieved substantially lower errors (MAE = 0.24, MRE = 4.46%) compared with the entropy-based approach (MAE = 2.72, MRE = 50.3%). Fig. 9 presents histogram of $\varepsilon_{r,\text{medium}}$ obtained using different AFT criteria and the $\varepsilon_{r,\text{RMS}}$ reference throughout the entire dataset. The SSIM-based AFT produces a distribution closely aligned with the theoretical range, whereas the entropy-based criterion systematically underestimates the permittivity by favoring lower values. This observation is consistent with the qualitative results in Fig. 8, indicating that the SSIM-based AFT more effectively diffuses reflection signals to the true defect boundaries even across large-scale validation cases. Overall, these results demonstrate that the proposed SSIM-based AFT yields a more robust, physically consistent estimate of effective permittivity than the conventional entropy-based AFT.

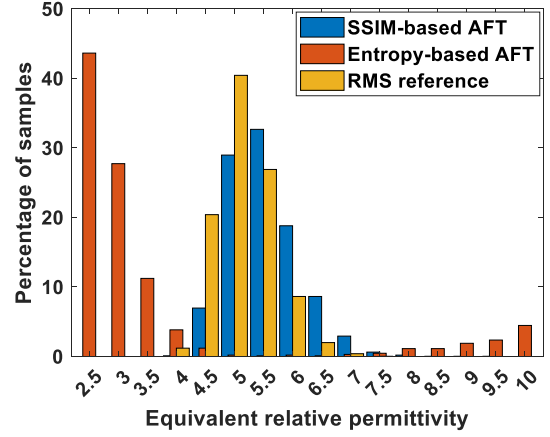

Fig. 9. Histogram of $\varepsilon_{r,\text{medium}}$ distributions across the entire synthetic dataset, estimated using the SSIM-based AFT, entropy-based AFT, and the RMS reference.

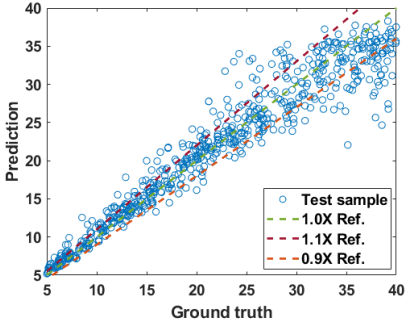

Fig. 10. Predictions of $\varepsilon_{r,\text{defect}}$ over the testing dataset.

### D. Results of DPE-Net

To evaluate the performance of the DPE-Net intuitively, both $\varepsilon_{r,\text{medium}}$ and $\varepsilon_{r,\text{defect}}$, which are normalized to [0,1] in the training process, are scaled back to their original value range. The DPE-Net with the best accuracy achieves an MAE of 1.95 and 0.25 for $\varepsilon_{r,\text{defect}}$ and $\varepsilon_{r,\text{medium}}$, respectively, and the corresponding MREs are 8.50% and 4.68%, respectively, demonstrating its capability of extracting permittivity values from the B-scan. Fig. 10 shows the prediction of $\varepsilon_{r,\text{defect}}$ over the entire testing dataset. It can be seen that $\varepsilon_{r,\text{defect}}$ of the majority of samples could be predicted within the tolerance of an MRE of 10%. The degraded predictions of samples located outside the margin of the reference lines are attributed to the loss of correlation between the intensity of the reflected signal and the relative permittivity of the defect when the EM signals penetrate into and propagate along the layered host medium. In general, all these performances demonstrate that the proposed DPE-Net successfully extracts the features of signals reflected from both the layered medium and the defect, and accurately decodes the key features to the corresponding permittivity values.

### E. Results of SR-Net

The performance of the defect's geometry reconstruction is evaluated by feeding the migrated image into the SR-Net. It should be noted that such migrated images are generated using the $\varepsilon_{r,\text{medium}}$ predicted from the DPE-Net. The SR-Net achieves an MSE of 0.0042, an SSIM of 0.9639, and an IoU of 0.8933. The low MSE demonstrates that SR-Net successfully recognizes and eliminates clutter and noise in the migrated images. The high SSIM and IoU not only reveal an accurate shape reconstruction of the internal defect but also indicate the precise location of the reconstructed defect inside the cylindrical medium. All these high-performance metrics demonstrate the robustness of the proposed SR-Net in recovering the correct geometrical information of defects.

TABLE I
ABLATION STUDY OF DPE-NET

| Model | A | B | C |
|---|---|---|---|
| FAM | × | ✓ | ✓ |
| CBAM | ✓ | × | ✓ |
| $MAE_{\text{defect}} \downarrow$ | 2.11 | 2.09 | **1.95** |
| $MAE_{\text{medium}} \downarrow$ | 0.27 | 0.26 | **0.25** |
| $MRE_{\text{defect}}$ (%) $\downarrow$ | 9.72 | 9.44 | **8.50** |
| $MRE_{\text{medium}}$ (%) $\downarrow$ | 4.90 | 4.72 | **4.68** |

TABLE II
ABLATION STUDY OF SR-NET

| Model | D | E | F |
|---|---|---|---|
| ResPath | × | ✓ | ✓ |
| CBAM | ✓ | × | ✓ |
| $MSE \downarrow$ | 0.0046 | 0.0046 | **0.0042** |
| $IoU \uparrow$ | 0.8865 | 0.8861 | **0.8933** |
| $SSIM \uparrow$ | 0.9623 | 0.9607 | **0.9639** |

### F. Ablation Studies of DPE-Net and SR-Net

Ablation studies are conducted for DPE-Net and SR-Net, in which the key modules of each network are systematically removed to analyze their particular contribution to the network's performance. Except for the ablated part, all the components of the networks are kept the same, and the ablated models are retrained accordingly with the same training parameters. Specifically, the contribution from the FAM and the CBAM are evaluated for the DPE-Net [Table I], while the effectiveness of the ResPath and the CBAM in the decoder are demonstrated for the SR-Net [Table II].

In Table I, the ablated models (Model A without FAM and Model B without CBAM) show reduced prediction accuracy across all metrics compared to the proposed DPE-Net (Model C). Similarly, Table II shows that removing the ResPath (Model D) or the CBAM in the decoder (Model E) from the SR-Net (Model F) leads to degraded performance. In general, the ablation studies of DPE-Net and SR-Net show the necessity and effectiveness of the modules in the designed networks.

### G. Visual Explanations of Model Decisions

To improve the interpretability of the proposed networks, gradient-weighted class activation mapping (Grad-CAM) [42] is employed to visualize the input regions most influential to the networks' predictions. Grad-CAM is applied to shallow convolutional layers of DPE-Net and SR-Net to retain higher spatial resolution and maintain alignment with the input. Fig. 11 shows two representative testing samples. For DPE-Net, the Grad-CAM heatmaps highlight distinct reflection patterns: in Fig. 11(b-1) and (b-2), the network focuses on reflections from tree trunk layers when estimating $\varepsilon_{r,\text{medium}}$, while in Fig. 11(c-1) and (c-2), the activations concentrate on reflections from decay regions when predicting $\varepsilon_{r,\text{defect}}$. These results confirm that DPE-Net learns to associate specific reflection responses in the B-scan with the corresponding dielectric properties. It is worth noting that the highlighted zones in the Grad-CAM visualizations indicate regions where the output is most sensitive, rather than representing the complete set of features that the model relies on for prediction. Since DPE-Net performs a global regression task, its estimates are governed by the overall statistical behavior of the entire B-scan.

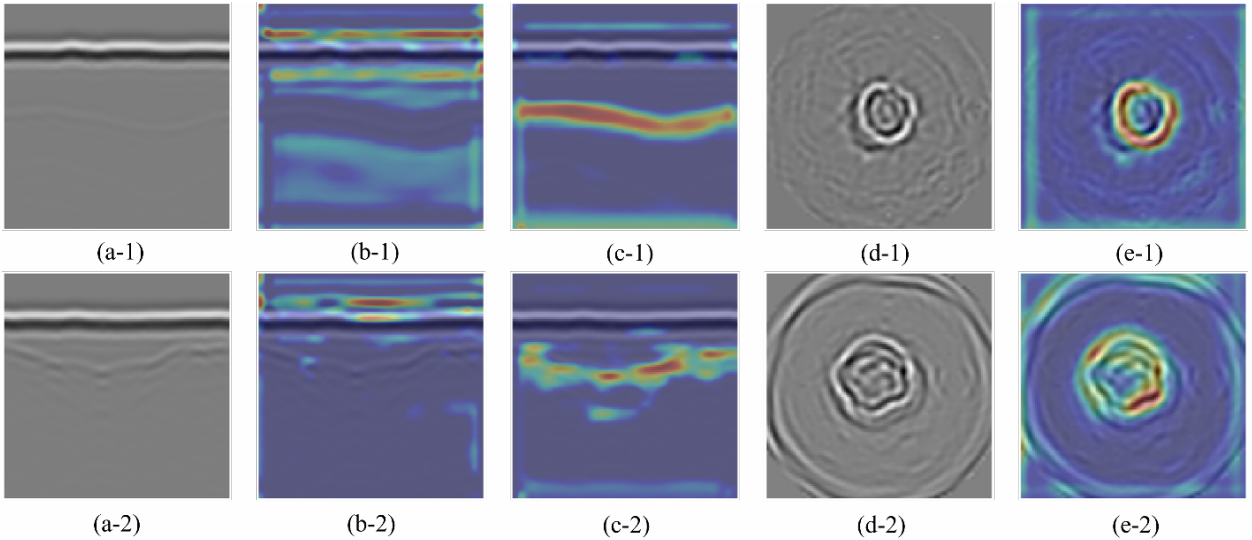

Fig. 11. Grad-CAM visualization on testing samples: (a-n) input B-scans for DPE-Net, (b-n) activation maps for predicting $\varepsilon_{r,\text{medium}}$, (c-n) activation maps for predicting $\varepsilon_{r,\text{defect}}$, (d-n) input migration images for SR-Net, and (e-n) activation maps for defect geometry reconstruction (n = 1,2).

For SR-Net, the Grad-CAM maps [Fig. 11(e-1) and (e-2)] consistently highlight the boundaries of the defect regions in the migrated images [Fig. 11(d-1) and (d-2)], indicating reliance on edge-related features to delineate defect structures. Weak background activations further show effective suppression of clutter and noise, ensuring predictions are guided by meaningful structural cues. Overall, these results demonstrate that both DPE-Net and SR-Net base decisions on physically and structurally meaningful features, reinforcing confidence in their applicability.

### H. Results and Comparison with Existing Deep Learning Schemes

The performance of the proposed scheme is demonstrated by comparing it to existing deep learning schemes, which are implemented using two advanced B-scan-based deep learning schemes: PiNet [32] and MRF-UNet [31], as well as one hybrid deep learning scheme, MultiPath-Net [33]. PiNet and MRF-UNet predict the permittivity map of the internal defect directly from the raw B-scan without incorporating an intermediate transformation. Due to the challenges of obtaining clutter-free B-scans of cylindrical objects, such as tree trunks, in real-world scenarios, the original MultiPath-Net is modified by removing its clutter-free B-scan encoding path. The revised network retains two encoding paths: one for the raw B-scan and the other for the migrated image determined by the entropy-based AFT. The permittivity map reconstructed by the proposed scheme is generated by combining the output from the DPE-Net and the SR-Net. For consistency, all models were trained and evaluated on the same dataset using identical preprocessing, initialization, and training configurations, detailed in Section III.B.

As shown in Table III, the proposed scheme achieves the best performance among all evaluated deep learning models. The relatively poor performances of PiNet and MRF-UNet highlight the limitations of B-scan-based approaches in reconstructing permittivity maps of defects inside cylindrical objects. The deterioration is primarily attributed to the absence of spatial alignment between the input B-scan and the output permittivity map, as explained in Section I via Fig. 1. Specifically, although MRF-UNet deploys a multi-receptive field module to enhance multi-scale feature representation, it fails to efficiently model the cross-domain transformation. PiNet introduces a global feature encoder to mitigate this issue, but its performance remains suboptimal. The hybrid MultiPath-Net performs worse than the proposed scheme due to two main factors: 1) The entropy-based AFT used to generate the migrated image has limited capacity to preserve defect shape accurately. 2) The simultaneous use of the raw B-scan and migrated image as inputs introduces confusion of feature representations, hindering robust decoding and ultimately impairing overall performance.

TABLE III
COMPARISON OF PERFORMANCE OF PROPOSED SCHEME WITH
EXISTING DEEP LEARNING SCHEMES USING SYNTHETIC DATASET

|  | MSE ↓ | SSIM ↑ | IoU ↑ |
|---|---|---|---|
| PiNet | 0.0039 | 0.9223 | 0.7309 |
| MultiPath-Net | 0.0042 | 0.9148 | 0.7352 |
| MRF-UNet | 0.0093 | 0.8272 | 0.3560 |
| Proposed | **0.0016** | **0.9634** | **0.8906** |

Qualitative comparisons, as shown in Fig. 12, further support these findings. The MRF-UNet yields inaccurate reconstructions [Fig. 12(c)], reflecting its difficulty in learning cross-domain transformation without spatial alignment. While the PiNet benefits from the global feature encoder and shows moderate improvement, the predictions of defects' shapes and permittivity [Fig. 12(b)] are not as satisfactory as the proposed scheme. MultiPath-Net [Fig. 12(d)] suffers from uneven reconstruction of the defect's permittivity map, which reveals the confusion of the decoding module in distinguishing and interpreting features from different domains of input. In contrast, the proposed scheme [Fig. 12(e)] delivers accurate and consistent reconstruction across diverse defects with both geometric structure and dielectric values being preserved, which affirms the method's robustness in reconstructing defects' permittivity maps.

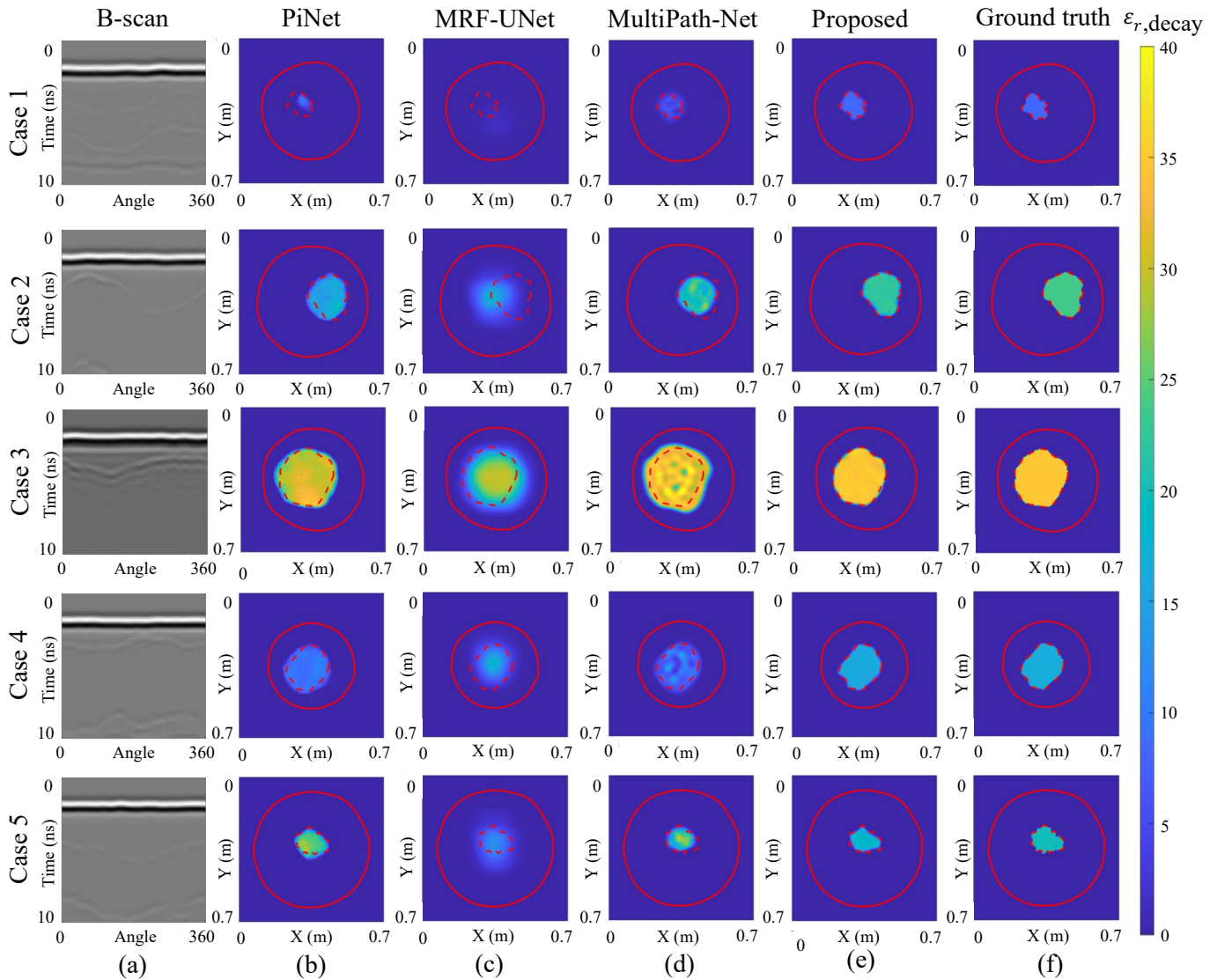

Fig. 12. (a) B-scans of five synthetic testing samples and the corresponding permittivity reconstruction results obtained using (b) PiNet, (c) MRF-UNet, (d) MultiPath-Net, and (e) the proposed scheme. (f) Ground truth. Red solid and dashed lines indicate the boundaries of the synthetic trunks and defects, respectively.

## IV. TESTS WITH REAL MEASUREMENT DATA OF LABORATORY TRUNK MODEL

### A. Dataset Generation and Implementation Details

The performance of the proposed scheme is further evaluated using measurement data from a laboratory tree trunk model, acquired via a stepped-frequency continuous-wave (SFCW) radar system that is specifically developed for tree trunk inspections [8]. As shown in Fig. 13(a), the system comprises a laptop, a vector network analyzer (Keysight VNA P5021A), and a high-gain, ultra-wideband, dual-polarized Vivaldi antenna with narrow beamwidth, operating between 0.5 and 3 GHz. The antenna provides sufficient penetration depth for the tree trunks investigated in this study and has been experimentally validated on the same tree species in our previous work [36]. The laboratory model is a cylindrical bucket filled with sand, selected for its dielectric properties similar to those of tree wood. To mimic internal decays with varying permittivity, four soil samples with different moisture contents—having relative permittivity values of 4, 8, 20, and 40 as measured by the Keysight N1501A Dielectric Probe Kit [Fig. 13(b)] —are prepared. The influence of decay shape is modeled using five geometries: circle, square, rectangle, triangle, and trapezoid. Each soil sample is divided into three portions to form decays of three different shapes, yielding 12 decay samples in total [Fig. 13(c)]. To further enhance dataset diversity, each decay sample is buried at 10 different locations within the laboratory trunk model, resulting in a total of 120 samples encompassing a wide range of defect shapes, sizes, positions, and permittivity values. Circular scanning is performed by mounting the model on an automated rotating platform. At every 6-degree increment, the radar system transmits electromagnetic waves and records an A-scan from the trunk model. A total of 60 A-scans are collected and stacked sequentially to form a B-scan of the object. Data processing plays a critical role in suppressing clutter and extracting defect signatures from raw B-scans. In this study, free-space removal is first applied to suppress the antenna's internal reflections. A Kaiser-window FIR filter centered at 1.5 GHz is subsequently applied to enhance the low-frequency components with better penetration capability, while suppressing high-frequency fluctuations arising from small structural irregularities or sharp localized reflections such as those produced by knots, resulting in B-scan input for the DPE-Net. For B-scans prepared for migration, an additional step using singular value decomposition is introduced to suppress layer-induced clutter and further highlight defect-related signatures [43]. Considering the natural variability of tree species, moisture conditions, and trunk geometries, these processing parameters may require adjustment when applied to other tree types or operational environments. The B-scan input into the DPE-Net is normalized to the [0, 1] range and resized to a resolution of 128 × 128 pixels. Since the permittivity values of the decay samples are highly unevenly distributed ({4, 8, 20, 40}), a log-scale min–max normalization is applied to map the labels to [0, 1], producing a more balanced representation across the permittivity range and enabling the model to learn uniformly from both low- and high-value samples [44]. The ground truth values of $\varepsilon_{r,\text{medium}}$ in the proposed scheme are determined by executing the migration algorithm over the permittivity range of [2, 3] with a step of 0.25 and selecting through the proposed SSIM-based AFT.

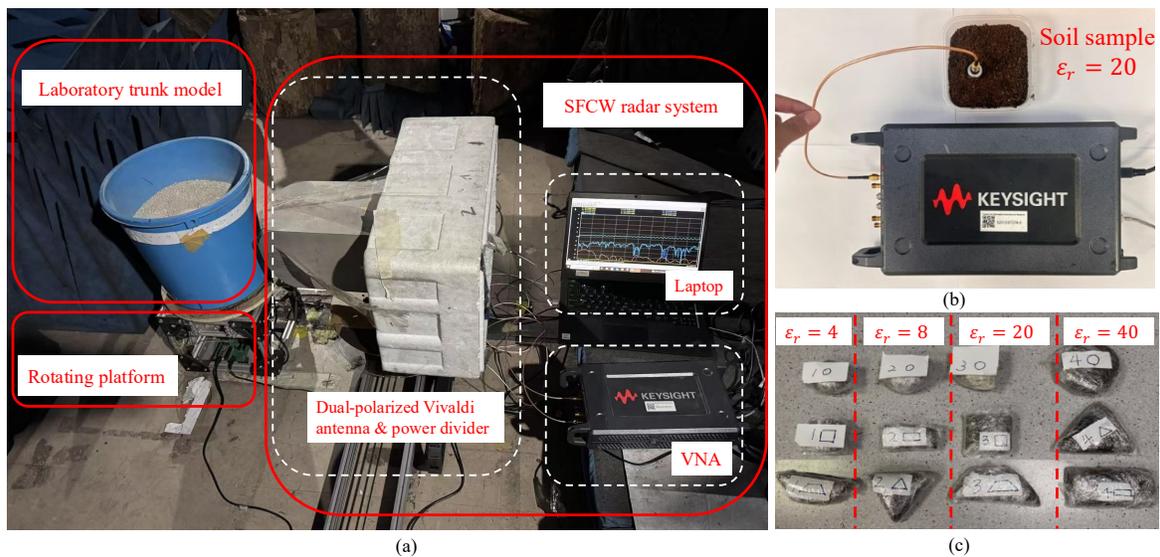

Fig. 13. (a) Laboratory trunk model and SFCW system, (b) permittivity measurement, and (c) soil samples with diverse shapes and permittivity

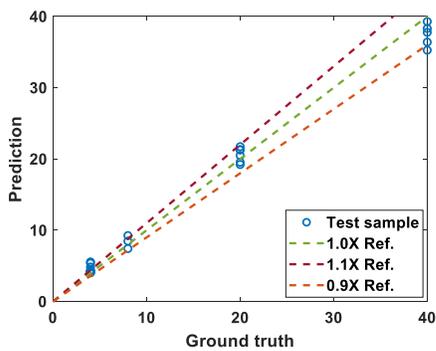

Fig. 14. Predictions of $\varepsilon_{r,\text{defect}}$ of testing samples in the measurement dataset of laboratory trunk model.

Once the dataset is generated, 80% of the data is randomly selected to fine-tune the scheme pre-trained by the synthetic data in Section III, and the remaining 20% is utilized for performance evaluation. All training configurations are kept consistent with those in Section III.B, except that the learning rate of the DPE-Net is reduced to 1e-5 for effective transfer learning. It should be noted that the predicted $\varepsilon_{r,\text{medium}}$ is used in the migration process and the resulting migrated image is fed to the SR-Net in the testing phase for a reasonable evaluation.

### B. Results and Comparison with Existing Deep Learning Schemes

The proposed DPE-Net achieves an MRE of 6.01% and an MAE of 0.14 for the estimation of $\varepsilon_{r,\text{medium}}$, and an MRE of 9.12% and an MAE of 1.16 for the prediction of $\varepsilon_{r,\text{defect}}$. As shown in

TABLE IV
PERFORMANCE COMPARISON OF THE PROPOSED SCHEME WITH
EXISTING DEEP LEARNING SCHEMES USING MEASUREMENT
DATASET OF LABORATORY TRUNK MODEL

|  | $MSE \downarrow$ | $SSIM \uparrow$ | $IoU \uparrow$ |
|---|---|---|---|
| PiNet | 0.0104 | 0.8590 | 0.7305 |
| MultiPath-Net | 0.0129 | 0.8659 | 0.7583 |
| MRF-UNet | 0.0124 | 0.8640 | 0.7083 |
| Proposed | **0.0104** | **0.9014** | **0.7947** |

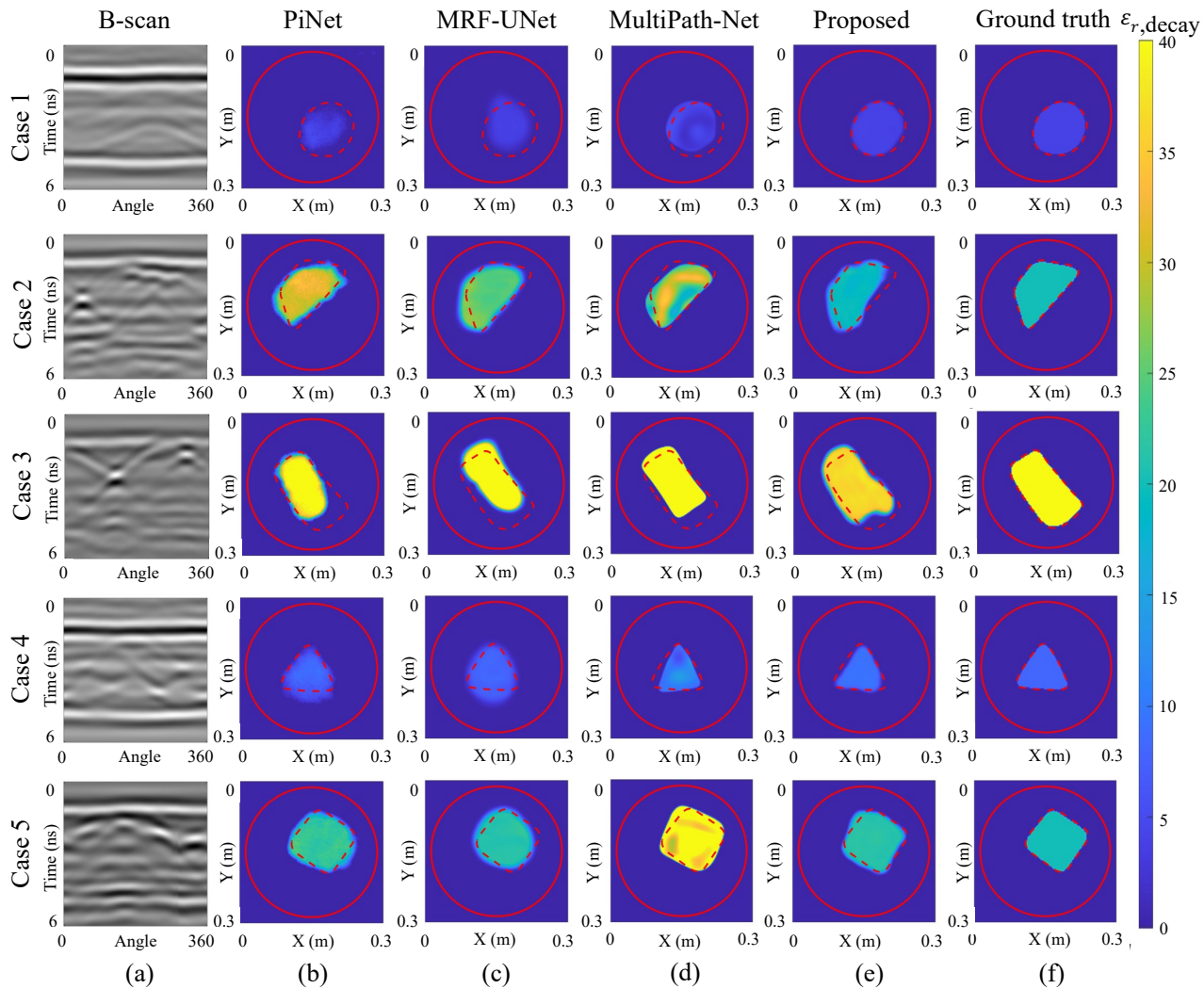

Fig. 15. (a) B-scans of five testing samples in the mesaurment dataset of laboraray tree trunk model and the corresponding permittivity reconstruction results obtained using (b) PiNet, (c) MRF-UNet, (d) MultiPath-Net, and (e) the proposed scheme. (f) Ground truth. Red solid and dashed lines indicate the boundaries of the synthetic trunks and defects, respectively.

Fig. 14, most of the predicted $\varepsilon_{r,\text{defect}}$ values fall within 10% of the ground truth, indicating the reliability of the DPE-Net for the measurement data. Given the limited size of the measurement dataset, performance could be enhanced with additional training samples.

The robustness of the proposed scheme is further proved by the quantitative results in Table IV, which shows its outperformance in all metrics. Although incorporating the migration image as an additional input, MultiPath-Net still exhibits poor IoU, suggesting confusion between features extracted from the B-scan and migration domains. Meanwhile, the degraded performances of PiNet and MRF-UNet highlight the intrinsic difficulty of directly mapping B-scans to permittivity images using real measurement data. Fig. 15 presents qualitative comparisons of reconstructed permittivity maps for five decay samples. In addition to accurately restoring the geometric and dielectric characteristics of circular defects, the proposed scheme effectively reconstructs irregular defects with sharp corners—particularly those with triangular, trapezoidal, and square shapes. In contrast, PiNet and MRF-UNet produce blurred boundaries and distorted shapes, indicating their limited cross-domain interpretability. Although MultiPath-Net, benefitting from the inclusion of migration images, can capture the overall shape of the defects, its erroneous orientations in triangular and trapezoidal cases, along with uneven reconstruction quality, reflect its difficulty in handling multi-domain features.

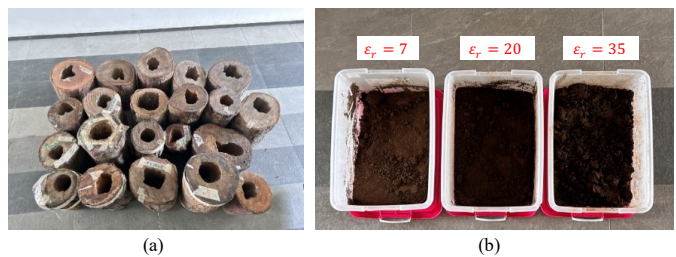

Fig. 16. (a) Fresh-cut and drilled tree trunk samples and (b) decay samples with different permittivity values.

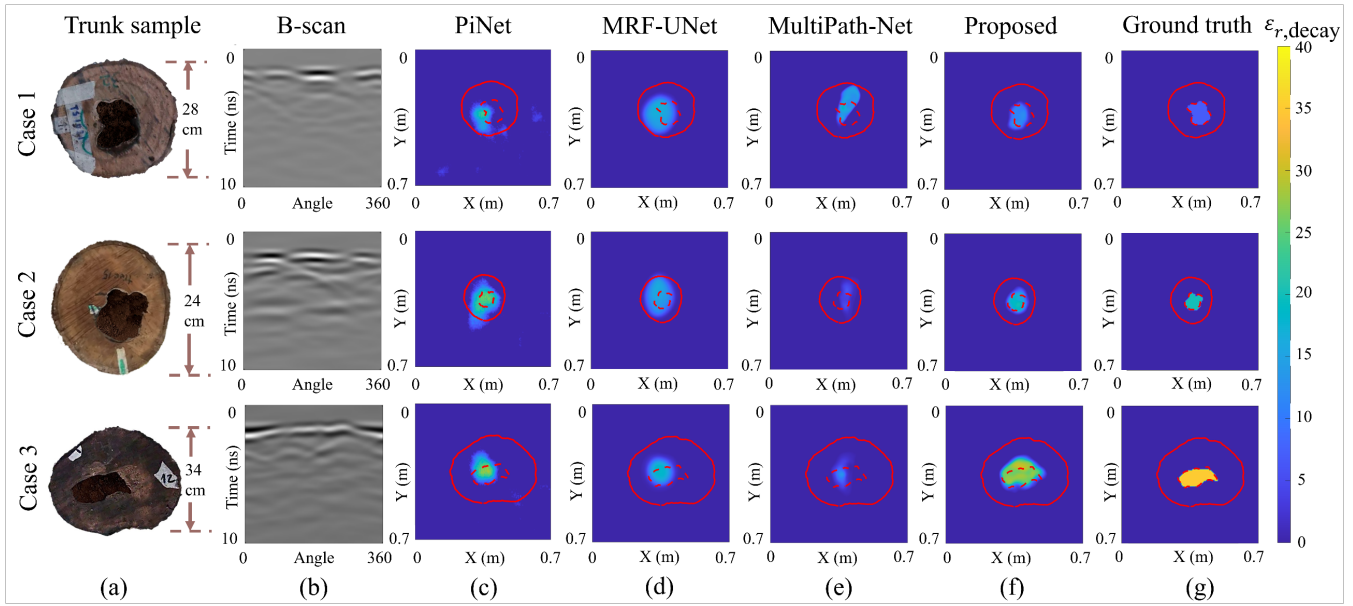

Fig. 17. Comparison of permittivity reconstruction results for three testing examples from the measurement dataset of real tree trunk samples: (a) photographs of the trunks, (b) their corresponding B-scans, (c–f) reconstructions obtained using PiNet, MRF-UNet, MultiPath-Net, and the proposed scheme, respectively, and (g) the ground truth. Red solid and dashed lines indicate the boundaries of the synthetic trunks and defects, respectively.

## V. TESTS WITH REAL MEASUREMENT DATA OF REAL TREE TRUNK SAMPLES

### A. Dataset Generation and Implementation Details

The capability of the proposed scheme is further validated using a measurement dataset of real tree trunk samples with artificially simulated defective regions. Twenty-one fresh-cut tree trunks representing the most common species in Singapore, including *Pterocarpus indicus*, *Tabebuia rosea*, and *Syzygium grande*, are collected [Fig. 16(a)]. The diameters and heights of these tree trunks range from [25, 50] cm and [35, 60] cm, respectively. Due to the difficulty of obtaining naturally decayed trunks, the defect regions in the dataset are simulated by drilling holes of arbitrary shapes and locations into each trunk and filling them with synthesized decay materials. The smallest size of the synthesized defect region is set to 8 cm, which is constrained by the range resolution of the deployed SFCW system and validated in our previous study [36]. The external trunk geometry and the shapes of the drilled holes are extracted from the cross-section images via edge detection, with the latter serving as the ground-truth defect shapes in our experimental setup. To represent different degrees of decay, three types of decay materials are synthesized by mixing soil samples with varying water contents, yielding permittivity values of [7, 20, 35] [Fig. 16(b)]. These values are measured using the Keysight N1501A Dielectric Probe Kit. Each hole was then filled with one of the measured materials, whose permittivity value served as the ground-truth permittivity of the corresponding defect region. Using the measurement setup and signal processing pipeline described in Section IV.A, a dataset comprising twenty-one processed B-scans is obtained. Ground-truth values of $\varepsilon_{r,\text{medium}}$ for DPE-Net are generated following the same procedure outlined in Section IV.A over [2, 4] with a step of 0.5.

To mitigate bias from the limited dataset size, seven-fold cross-validation is adopted. Specifically, twenty-one tree trunks are randomly divided into seven folds. Each fold contains three individual tree trunks with distinct decay materials. In each iteration, six folds are used for training, and the remaining fold is used for testing, yielding seven models trained on different fold combinations. The learning rates for DPE-Net and SR-Net are set to 1e-5 and 1e-6, respectively, while all other training configurations remain the same as described in Section III.B.

### B. Results and Comparison with Existing Deep Learning Schemes

Table V presents the quantitative comparison between the proposed scheme and the other schemes. To improve statistical confidence, both the mean and standard deviation of each metric are computed over the seven-fold cross-validation, covering the model's performance on every sample in the measurement dataset. The proposed scheme achieves the best averaged performance among all deep learning models with satisfactory standard deviations, indicating its robust performance across different tree trunk samples. Most importantly, the notably superior IoU metric highlights its capability to localize and estimate defect regions, demonstrating the effectiveness of incorporating the physics-informed migration algorithm into the deep learning framework. The effectiveness of the proposed scheme is further illustrated in Fig. 17, where the reconstructed permittivity maps of three examples in the testing dataset are shown. PiNet and MRF-UNet reconstruct the internal decays as blur regions with imprecise shapes and inaccurate permittivity values, reflecting the difficulty of B-scan-based schemes in handling domain transformation without explicit spatial alignment. While MultiPath-Net performs better than these two baselines in the first example, its performance degrades in the remaining cases. The low IoU value further confirms its confusion when interpreting the encoded features from two different domains. In contrast, the proposed migration-assisted deep learning scheme achieves superior reconstructions across all three examples. It accurately recovers not only the locations of the defects but also their approximate shapes, sizes, and aspect ratios, closely matching the ground truth. Notably, these results are obtained using only 18 measurements for fine-tuning the pre-trained scheme, suggesting that its performance could be further improved with a larger dataset.

TABLE V
COMPARISON OF PERFORMANCE OF PROPOSED SCHEME WITH EXISTING DEEP LEARNING SCHEMES USING MEASUREMENT DATASET OF REAL TREE TRUNK SAMPLES

| | MSE ↓ (× 10⁻³) | | SSIM ↑ (× 10⁻²) | | IoU ↑ (× 10⁻²) | |
|---|---|---|---|---|---|---|
| | Ave. | Std. | Ave. | Std. | Ave. | Std. |
| PiNet | 6.0 | 6.6 | 92.11 | 2.66 | 40.01 | 20.30 |
| MultiPath-Net | 6.9 | 7.2 | 94.01 | 1.57 | 14.85 | 20.54 |
| MRF-UNet | 4.8 | **4.7** | 93.06 | **0.73** | 44.38 | 23.55 |
| Proposed | **4.8** | 5.0 | **94.80** | 1.89 | **54.74** | **16.12** |

TABLE VI
COMPUTATIONAL TIME COMPARISON ON A SINGLE TEST SAMPLE

|  | PiNet | MRF-UNet | MultiPath-Net | | | Proposed | | | |
|---|---|---|---|---|---|---|---|---|---|
|  |  |  | Migration | Network | Total | DPE-Net | Migration | SR-Net | Total |
| CPU Time(s) | 2.07 | 3.70 | 87.73 | 2.08 | 89.81 | 0.18 | 17.62 | 2.24 | 20.04 |

TABLE VII
ZERO-SHOT PERFORMANCE COMPARISON ON REAL TREE TRUNK
MEASUREMENT DATA

|  | $MSE \downarrow$ $(\times 10^{-3})$ | | $SSIM \uparrow$ $(\times 10^{-2})$ | | $IoU \uparrow$ $(\times 10^{-2})$ | |
|---|---|---|---|---|---|---|
|  | Ave. | Std. | Ave. | Std. | Ave. | Std. |
| PiNet | 0.74 | 0.85 | 93.83 | 1.36 | 6.25 | 9.32 |
| MultiPath-Net | 5.27 | 2.34 | 83.74 | 3.44 | 27.26 | 10.61 |
| MRF-UNet | 1.43 | 1.07 | 75.87 | 3.73 | 24.78 | 16.47 |
| Proposed | **0.62** | **0.60** | **93.37** | **1.98** | **38.72** | **19.04** |

To assess the suitability of each method for on-field tree inspection, all methods are executed on a Dell Inspiron 16 laptop equipped with a 13th Gen Intel(R) Core(TM) i7-1360P processor (2.20 GHz) and 16 GB RAM to evaluate the CPU time. As presented in Table VI, PiNet and MRF-UNet achieve the shortest CPU times of 2.07 s and 3.70 s, respectively, as they directly transform the input B-scan into the corresponding permittivity map without any physical modeling. In contrast, MultiPath-Net requires multiple executions of the migration algorithm to determine the optimal migrated image using the entropy-based AFT, resulting in significantly higher computational time. The proposed scheme attains a more favorable trade-off between accuracy and efficiency. Specifically, DPE-Net predicts $\varepsilon_{r,medium}$ within one second, allowing the migration algorithm to be executed only once. Consequently, the total CPU time of the proposed framework is approximately 20 s, making it well-suited for rapid on-site reconstruction of tree defect permittivity maps.

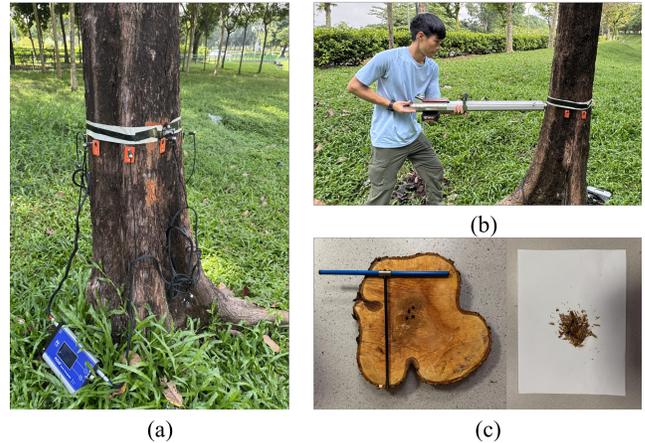

Fig. 18. Validation of the trees used in the generalization tests using (a) sonic tomography and (b) drilling resistance tomography, followed by (c) extraction of defective wood samples from confirmed regions using an incremental borer.

*C. Zero-Shot Tests*

To further evaluate the generalization capability of the proposed model and verify the effectiveness of simulation-based pre-training, zero-shot experiments are conducted on the measurement dataset of real tree trunk samples. In this setting, the models trained solely on the synthetic dataset described in Section III are directly applied to the real measurements without any fine-tuning. As summarized in Table VII, the proposed scheme achieves the best overall performance among all compared deep learning models, demonstrating its strong capability of transferring knowledge learned from simulation to real measurement conditions. Benefiting from the migration-assisted scheme design, the proposed method effectively mitigates spatial misalignment between the input B-scan domain and the output imaging domain, resulting in a notably higher IoU and thus more accurate defect localization. These results confirm that the pre-training stage enables the model to capture generalized and physically meaningful defect-related representations that remain valid in real-world scenarios. It should be noted that the comparison between Table V and Table VII indicates that fine-tuning on measured data further enhances the model's performance. This improvement confirms that transfer learning effectively bridges the domain gap between simulated and real GPR data, enabling the model to suppress unwanted clutter in the B-scans of real tree trunks and to better capture the underlying physical and statistical characteristics of real-world measurements.

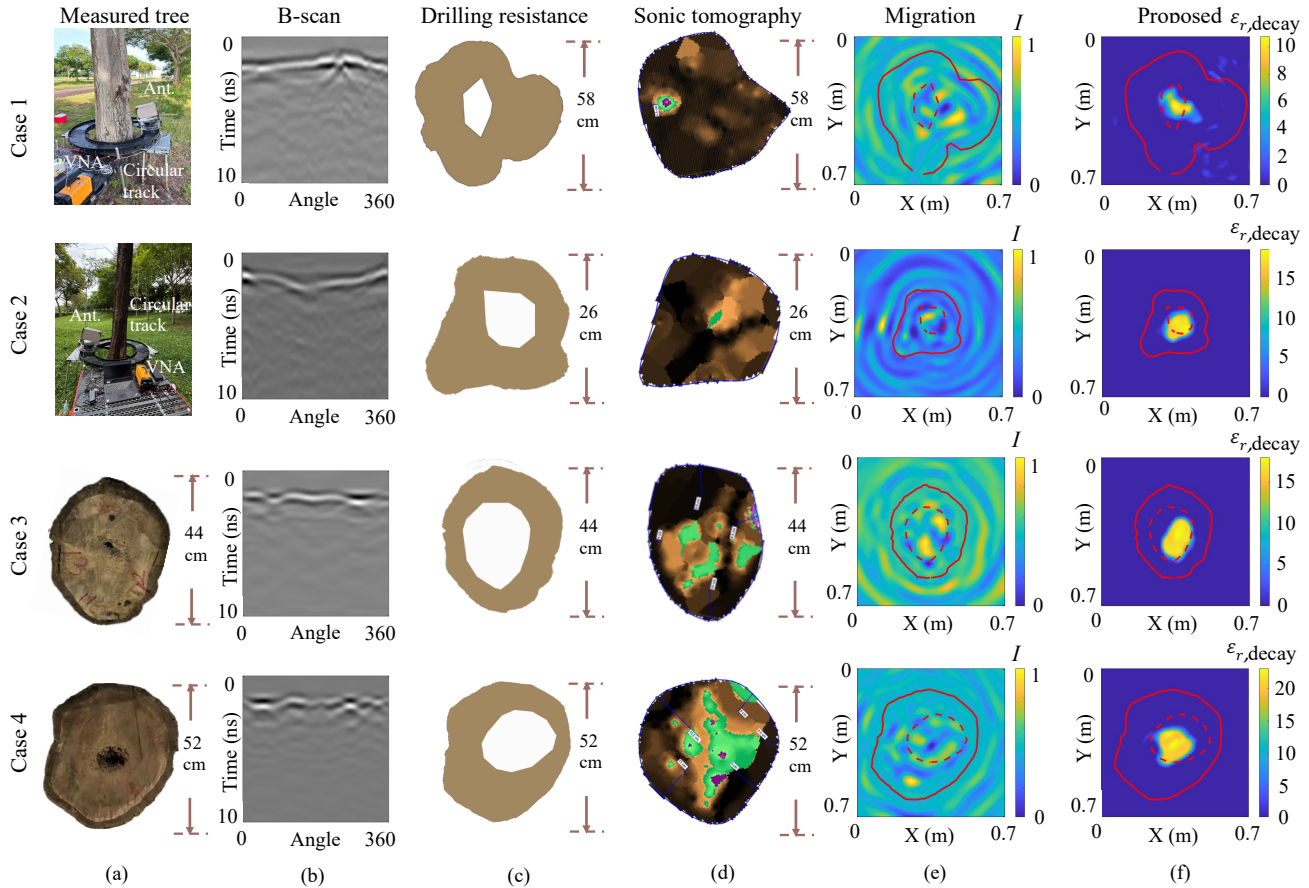

Fig. 19. (a) Real trees used in the generalization tests and their corresponding (b) B-scans. Reconstruction results obtained by (c) drilling resistance tomography, (d) sonic tomography, (e) conventional imaging algorithms, and (f) the proposed method.

## VI. GENERALIZATION TESTS ON TREE TRUNKS WITH NATURALLY DEVELOPED DEFECTS

To evaluate the applicability of the proposed scheme under real-world conditions, generalization tests are conducted on tree trunks with naturally developed internal defects, including two freshly cut samples and two live trees. Scanning of the fresh-cut trunks is performed under the same laboratory conditions described in Section IV, within two hours after felling, to preserve moisture and structural integrity comparable to those of living trees. The tests of live trees are conducted on defective trees identified by arborists using invasive drilling resistance method and non-invasive sonic tomography [Fig. 18].

### A. Measurement Configurations

A customized circular slider system is developed for circumferential scanning of live trees [Fig. 19(a) (Case 1 and Case 2)]. The system consists of a modular circular track, assembled from interlocking 3D-printed base segments, and a motor-driven antenna platform controlled by an Arduino unit. Two flexible belts attached along the inner and outer track edges define its curvature and allow adaptation to different trunk diameters. During operation, the track is mounted on portable tables to encircle the tree at a fixed height, and the antenna platform automatically traverses the circular path while maintaining a constant radial orientation toward the trunk center. A-scans are sequentially acquired at predefined angular intervals, and upon completing one full scanning, they are combined to generate a B-scan of the tree trunk cross-section.

The geometry of the defective regions is estimated using a drilling-resistance measurement system (IML-RESI PowerDrill) and is used as the reference standard for evaluating the reconstruction results. The device drives a fine needle through the trunk from multiple sides to obtain one-dimensional resistance profiles that indicate decayed zones along each drilling path [Fig. 18(b)]. These profiles are combined to reconstruct a two-dimensional tomographic map approximating the defect shape. Additionally, ultrasonic tomography is performed prior to felling using the PiCUS Sonic Tomograph: sensors mounted circumferentially around the trunk recorded sonic-wave travel times, which are processed to reconstruct internal tomograms for comparison with the proposed method [Fig. 18(a)]. For permittivity validation, wood cores are extracted from decayed regions after felling using an incremental borer; the extracted samples are flattened to ensure proper probe contact and their dielectric properties measured with the Keysight N1501A Dielectric Probe Kit [Fig. 18(c)]. The external geometry of the scanned trunk section is obtained by capturing a 3D point cloud using the lidar sensor, projecting

| | | Case 1 | Case 2 | Case 3 | Case 4 |
|---|---|---|---|---|---|
| $IoU\uparrow$ $(\times 10^{-2})$ | Sonic tomography | 0 | 11.48 | 16.08 | **41.07** |
| | Migration | 4.98 | 27.77 | 24.68 | 30.37 |
| | Proposed | **32.32** | **48.53** | **44.58** | 40.20 |
| Decay permittivity | Ground truth | 6.28 | 14.05 | 14.91 | 25.88 |
| | Prediction | 10.76 | 19.01 | 18.07 | 23.09 |

the points onto a 2D plane, and reconstructing their concave hull polygon using the α-shape theory.

*B. Results and Comparative Analysis with Existing Field-Deployable Methods*

A qualitative comparison among drilling resistance tomography, sonic tomography, the traditional migration algorithm [23], and the proposed framework is provided in Fig. 19. Taking the invasive drilling resistance profiles as the reference, sonic tomography is generally able to localize the defective regions, but the reconstructed boundaries are spatially expanded and considerably blurred. This limitation arises because sonic tomography reflects the reduction of mechanical stiffness rather than the true morphological boundaries of internal decay. Additionally, the inversion process aggregates wave responses from multiple anisotropic and heterogeneous propagation paths within wood, which further degrades the geometric fidelity of the reconstructed defects [45]. The traditional migration algorithm reveals strong scattering responses at defect locations; however, such responses are often spatially discontinuous and fail to accurately convey the extent and morphology of the defective regions. The presence of structural noise and clutter may introduce spurious reflections, potentially leading to misinterpretation and reduced diagnostic reliability.

In contrast, the proposed method consistently produces accurate defect localization across all test samples. Although minor discrepancies remain in the precise shape or size of the decay, the reconstructed results reliably capture the correct defect positions and overall profiles. Particularly in challenging cases (e.g., Case 1), where sonic tomography and the migration algorithm fail to resolve the decay structures, the proposed method still provides satisfactory and interpretable reconstructions. These observations collectively confirm that the proposed method not only achieves higher reconstruction accuracy on average, but also maintains stability under some challenging conditions. Furthermore, the IoU for each sample, summarized in Table VIII, quantitatively verify the capability of the proposed framework to recover both the location and geometry of internal defects in all test cases. These findings demonstrate the effectiveness of integrating deep-learning-based estimation with physics-informed imaging in a unified manner for reliable tree-trunk inspection. It is worth noting that slight intensity variations may appear within the reconstructed permittivity maps. This visual phenomenon arises because the SR-Net employs a ReLU activation in its final layer, producing continuous-valued geometry predictions near defect boundaries rather than strictly binary outputs. When these continuous values are combined with the defect permittivity estimated by the DPE-Net, gradual intensity changes may appear inside the defect region. In addition, the permittivity estimation inside defective regions may exhibit noticeable errors in several cases. This degradation reflects both the inherent difficulty of accurately predicting the permittivity of decayed wood in practical scenarios due to the biological variability among tree species, and the domain shift between synthesized training defects and the more complex morphological characteristics of natural decay. As future work, fine-tuning the network using measurement datasets acquired from live trees is expected to mitigate the domain gap and further enhance permittivity prediction reliability.

*C. Discussion and Future Work*

Despite demonstrating the feasibility of applying the proposed scheme to tree trunks with naturally developed defects in real-world conditions, several limitations should be acknowledged. First, due to the inherent difficulty of acquiring sufficient live-tree data, fine-tuning and validation are performed on a measurement dataset comprising a limited number of real tree trunk samples. While this dataset demonstrates the feasibility and outperformance of the proposed framework, its limited size may compromise the statistical robustness of the results. Second, the defective regions in the measurement dataset are artificially created by drilling cavities into freshly cut tree trunk samples and filling them with soil-based materials of controlled permittivity. This design may not fully reproduce the complexity of natural decay. Moreover, the moisture variation of the fresh-cut tree trunk sample during short storage may still alter the host medium's dielectric properties, leading to minor deviations from live-tree conditions. These factors degrade the scheme's performance in real applications. Third, the current study focuses on trunks with approximately convex and regular cross-sections. For species exhibiting irregular or non-convex geometries, surface undulations can cause multiple scattering between adjacent bark surfaces and spurious echoes that resemble internal defects. Such geometric irregularities lower the interpretability value of the B-scan, thereby challenging the accurate reconstruction of the defect. Finally, the validation is conducted on the most common tree species in Singapore. For trees with higher water content or thicker cambium layers, strong attenuation in the outer tissues may weaken or even eliminate reflections from internal defects, making GPR inspection difficult. Antenna designs with higher gain, narrower beams, and optimized frequency bands should therefore be considered to improve penetration and detection reliability.

Future work aims to address these limitations through both data expansion and methodological refinement. A larger and more diverse dataset of live-tree measurements will be collected using the proposed circular scanning system to capture interspecies geometric and moisture variability. In parallel, advanced signal processing techniques supported by digital-twin–based approaches will be developed to model and suppress multiple reflections and clutter induced by irregular tree trunk shapes. Finally, the deep-learning–assisted full-waveform inversion scheme will be further explored to enable more accurate permittivity estimation and a quantitative assessment of the comprehensive dielectric distribution inside tree trunks, thereby advancing GPR-based tree health diagnostics in real-world scenarios.

## VII. Conclusion

This study introduces a migration-assisted deep learning scheme for reconstructing the permittivity map of the defect in cylindrical objects. First, a DPE-Net is designed to extract features from B-scans and predict both the defect's permittivity and the host medium's equivalent permittivity. An SSIM-based AFT is introduced to determine the ground truth of the host medium's equivalent permittivity to train DPE-Net, thereby enabling the generation of a migrated image that accurately preserves the defect's geometric profile. Second, the modified Kirchhoff migration algorithm maps the reflected signals to the imaging domain using the predicted host medium's equivalent permittivity. Third, the migrated image is processed by an SR-Net, which incorporates ResPath and attention-based decoders to suppress the clutter, enhance structural clarity, and output a precise defect geometry map. Extensive validations are conducted using the synthetic layered cylinder dataset, the measurement datasets of a laboratory trunk model and real tree trunk samples. In all scenarios, the proposed method outperforms existing deep learning schemes, demonstrating its robustness. Generalization tests on live trees further verify that the proposed method reliably identifies both the size and spatial location of internal defects, highlighting its strong potential for practical tree inspection. While the framework is primarily developed for tree defect imaging, it is broadly applicable to cylindrical object inspection and can be extended to other GPR imaging contexts involving the circumferential scanning.

### Acknowledgement

The authors would like to thank the editors and anonymous reviewers for their valuable comments and suggestions. The authors also gratefully acknowledge Jonathan Goh and Nguyen Thanh Tin for their assistance in conducting the field tests.